\documentstyle[12pt]{article} 
\begin{document}
\baselineskip=20pt

\begin{center}
{\Large {\bf Normal ordering and boundary conditions for fermionic string coordinates}}\\
\vspace{1cm} 

{\large Nelson R. F. Braga$^{a}$ , Hector L. Carrion$^{b}$ and 
Cresus F. L. Godinho$^{c}$  } \\
 
\vspace{0.5cm}
{\sl
$^a$ Instituto de F\'{\i}sica, Universidade Federal
do Rio de Janeiro,\\
Caixa Postal 68528, 21941-972  Rio de Janeiro, RJ,  Brazil\\[1.5ex]

$^b$ Instituto de F\'{\i}sica, Universidade de S\~ao Paulo,\\
Caixa Postal 66318, 05315-970, S\~ao Paulo, SP, Brazil\\[1.5ex]

$^c$ Centro Brasileiro de Pesquisas F\'{\i}sicas, Rua Dr Xavier Sigaud 150,\\
 22290-180,
Rio de Janeiro, RJ, Brazil }
\end{center}
\vspace{1cm}

\abstract  
We build up normal ordered products for fermionic open string coordinates
consistent with boundary conditions.
The results are obtained considering the presence of antisymmetric tensor fields.
We find a discontinuity of the normal ordered products at string endpoints 
even in the absence of the background.
We discuss how the energy momentum tensor also changes at the world-sheet 
boundary in such a way that the central charge keeps the standard value at 
string end points.


\vspace{1cm}

\noindent braga@if.ufrj.br; hlc@fma.if.usp.br; godinho@cbpf.br

\vfill\eject


Recent progress in string theory\cite{Malda,RS1,RS2} indicates scenarios where our 
four dimensional space-time with standard model fields corresponds to a D3-brane\cite{Po} 
embedded in a larger manifold. 
Since D-branes correspond, in type II string theories, to the space where the open string 
endpoints are attached, our space-time would be affected by string boundary conditions.
One important consequence is the possible non-commutativity of space-time 
coordinates at very small length scales\cite{CH1,Alekseev:1999bs,SW}
since commuting coordinates are incompatible
with open string boundary conditions in the presence of antisymmetric tensor backgrounds. 
This is one of the reasons for the increasing interest
in many aspects of non-commutative quantum field theories
as can be seen for example in\cite{SW,REVIEW}.   
Furthermore, this illustrates the fact that string boundary 
conditions may play a non trivial role in our four dimensional physics.

In quantum field theory, products of quantum fields at the same space-time points 
are in general singular objects. The same thing happens in string theory with 
position operators, that can be taken as conformal fields on the world-sheet.
This situation is well known and one can remove the singular part of the 
operator products by defining normal ordered non-singular objects\cite{Po2}.
The explicit characterization of these singularities is important
when one investigates the realization, at quantum level,
of the classical symmetries. The singular terms that show up in products 
of energy momentum tensors are associated with the central charge
and play a crucial role in fixing the critical dimension where conformal invariance holds.

Normal ordered products of operators are usually defined so as to satisfy
the classical equations of motion at quantum level.
For open strings the position operators are defined in a manifold with boundary 
(the world-sheet). So they also have to satisfy boundary conditions.
For the bosonic sector of string coordinates, it was shown in \cite{Braga:2004wr}
how to build up normal ordered products of operators consistent with boundary 
conditions and equations of motion. In that bosonic case the effect of the boundary 
was to add extra terms to the normal ordering. This causes no change 
in the calculation of the central charge.

Here we calculate the normal ordered products for fermionic open string coordinates
in a flat embedding space with a constant antisymmetric tensor background, taking the boundary conditions into account.
We will see that the products of fermionic position operators 
of the same kind get a factor of $1/2$ at string endpoints.
This factor have to be considered in the calculation of 
the critical dimension for string theory, since normal ordering is used to 
build up the algebra of conformal transformations.
However we will see that the boundary conditions also affect the energy 
momentum tensor on the boundary in such a way that the central charge is unchanged.  


A superstring in Minkowiski space in the presence of a constant antisymmetric 
tensor field ${\cal B}^{\mu\nu}$ can be described by the action \cite{a,Lindstrom:2002mc,Braga:2003ic, Chakraborty:2005pt}
\begin{eqnarray} \label{a1}
 S &=&  \frac{- 1}{4 \pi \alpha^\prime} \int_{\Sigma} d\tau
 d{\rho}\,\Big[ \, \partial_a X^\mu \partial^a X_\mu 
\,+\, \epsilon^{ab} B_{\mu\nu} \partial_a X^\mu \partial_b X^\nu
\nonumber\\
 & & + i \psi_{\mu (-)} E^{\nu \mu} \partial_{+} \psi_{\nu (-)} +
i \psi_{\mu (+)} E^{\nu \mu} \partial_{-} \psi_{\nu (+)} \,\Big]
\end{eqnarray}
 
\noindent where $\partial_{+} =  \partial_{\tau} + \partial_{\rho},\; \;
\partial_{-} =  \partial_{\tau} - \partial_{\rho} $
 and $ E^{\mu\nu} \,=\, \eta^{\mu\nu} \, + {\cal B}^{\mu\nu}\,$.

In a flat embedding space, as is the case here, bosonic and fermionic string coordinates satisfy independent boundary conditions
(for string boundary conditions in general spaces see \cite{Albertsson:2001dv,Albertsson:2002qc}). 
It is interesting to remark that the coupling of the superstring to the antisymmetric tensor field background has to be defined in such a way that boundary conditions do not spoil supersymmetry.
For the action (\ref{a1}) it was shown in \cite{a,Lindstrom:2002mc} that boundary conditions are consistent with supersymmetry in the sense that the supersymmetry transformation of the fermionic boundary conditions leads to the bosonic boundary conditions. 

From now on we will just consider the fermionic sector of the string action since the bosonic one was already studied in \cite{Braga:2004wr}. 
The volume terms coming from the minimum action condition imply the classical equations 
of motion: 
\begin{equation}
\partial_{+} \psi_{\nu (-)} \,=\, 0\,\,\,,
\partial_{-} \psi_{\nu (+)} \,=\, 0\,.
\end{equation}

While the boundary term leads to the condition
\begin{equation}
\label{3.1}
\Big( \psi^\mu_{(-)} \,E_{\nu\mu}\,  \delta \psi^\nu_{(-)}\,-\,
\psi^\mu_{(+)} \, E_{\nu\mu} \, \delta \psi^\nu_{(+)}
\Big)\vert_{0}^\pi\,\,=\,\,0\,\,,
\end{equation}

\noindent that is satisfied imposing the constraints
\begin{equation}
\label{BC}
 E_{\nu\mu}\,  \psi^\nu_{(+)} (0,\tau)\,=\,
  E_{\mu\nu} \, \psi^\nu_{(-)} (0,\tau)\,
\end{equation}
\begin{equation}
\label{BC2}
E_{\nu\mu}\,  \psi^\nu_{(+)} (\pi,\tau ) \,=\,\lambda
E_{\mu\nu} \, \psi^\nu_{(-)} (\pi, \tau )\,\,,
\end{equation}

\noindent at the endpoints $\rho \,=\,0$ and $\rho = \pi\,$,
where $\lambda = \pm 1 \,$ corresponding to Ramond or Neveu-Schwarz boundary conditions.

Now in order to change to complex world-sheet coordinates we first take the Euclidean
form of the fermionic action by means of the transformation $\,\rho \,\,=\,\, -\imath\, \sigma\,\,$.
Then we introduce the complex variables: $z \,=\, \tau \,+ i \sigma \,\,,\,\,
{\bar z} \,=\,\tau \,- i \sigma \,$ and the action takes the form
\begin{eqnarray}
 S =  \frac{-\imath}{4 \pi \alpha^\prime} \int_{\Sigma} dz
 d \bar{z}[\psi_{\mu (-)} E^{\nu \mu} \partial_{\bar{z}} \psi_{\nu (-)} +
 \psi_{\mu (+)} E^{\nu \mu} \partial_{z} \psi_{\nu (+)}]\,\,.
\end{eqnarray}

\noindent In this complex coordinates the string end-points 
corresponds to the regionz $z \,=\,{\bar z}\,$ and $z \,=\,{\bar z}+ 2\,i\,\pi\,$.

As usual, one can study the properties of quantum operators by considering the 
corresponding expectation values, defined in terms of path integrals.  
Using the fact that the path integral of a 
total functional derivative vanishes and considering the insertion of one 
fermionic operator  one finds
\begin{equation}
\label{insertion}
\int [ d  \psi ] \left[  \frac{\delta}{\delta \psi^{\mu} _{(a) }(z,\bar{z})} 
[e^{-S} \psi^{\nu}_{(b)} (z', \bar{z}')]
  \right] \,=\,0\,\,,
\end{equation}

\noindent where $a,b \,=\,+,-\,$. Considering first the case of 
$\psi^{\nu}_{(b)} (z', \bar{z}')\,$ inside the world-sheet and not at the boundary. 
That means: $ z' \ne \bar{z}'\,$, this equation implies the following expectation values
\begin{eqnarray}
\label{eqmov}
< \partial_{z} \psi^{\mu}_{(+)} (z,\bar{z}) \psi^{\nu}_{(+)}
(z',\bar{z}') > &=& 2\,\pi \,i \, \alpha'\, < \eta^{\mu \nu} 
\delta^2(z-z',\bar{z}-\bar{z}') > \nonumber \\
< \partial_{\bar{z}} \psi^{\mu}_{(-)} (z,\bar{z})
\psi^{\nu}_{(-)}(z',\bar{z}') > &=& 2\,\pi \,i \, \alpha '\, < \eta^{\mu
\nu}  \delta^2(z-z',\bar{z}-\bar{z}') > \nonumber \\
< \partial_{\bar{z}} \psi^{\mu}_{(-)} (z,\bar{z})
\psi^{\nu}_{(+)}(z',\bar{z}') > &=&  < \partial_{z} \psi^{\mu}_{(+)} (z,\bar{z})
\psi^{\nu}_{(-)}(z',\bar{z}') > =  0\,\,.
\end{eqnarray}

From these results we find the appropriate way to 
define normal ordered products that satisfy the equations of motion
for fermionic operators that are not at the world-sheet boundary     
\begin{eqnarray}
\label{no}
: \psi^{\mu}_{(+)}(z,\bar{z})\,\,\, \psi^{\nu}_{(+)}(z',\bar{z}')  : &=& 
  \psi^{\mu}_{(+)}(z,\bar{z})\,\, \,\psi^{\nu}_{(+)}(z',\bar{z}') 
\,-\, \frac{i\, \alpha'\,}{{\bar z} - {\bar z}'}\,\eta^{\mu \nu}
\nonumber\\
: \psi^{\mu}_{(-)}(z,\bar{z})\,\,\, \psi^{\nu}_{(-)}(z',\bar{z}')  : &=& 
  \psi^{\mu}_{(-)}(z,\bar{z})\,\,\, \psi^{\nu}_{(-)}(z',\bar{z}') 
\,-\, \frac{i\, \alpha'\,}{ z -  z'}\,\eta^{\mu \nu}
\nonumber\\
: \psi^{\mu}_{(+)}(z,\bar{z})\,\,\, \psi^{\nu}_{(-)}(z',\bar{z}')  : &=& 0
\nonumber\\
: \psi^{\mu}_{(-)}(z,\bar{z})\,\,\, \psi^{\nu}_{(+)}(z',\bar{z}')  : &=& 0
\end{eqnarray}

These products satisfy the quantum equations of motion,
that in complex coordinates read  
$\partial_{z} \psi^{\mu}_{(+)}\,=\,0\,;\,\partial_{\bar{z}} \psi^{\mu}_{(-)} \,=\,0\,$),
as for example $ \partial_{\bar{z}}  : \psi^{\mu}_{(-)} (z,\bar{z}) 
\,\,\psi^{\nu}_{(-)}(z',\bar{z}') \,:\,=\,\,0\,\,\,$ 
and are not subject to boundary conditions since they are defined for points 
inside the world-sheet.

Let us now consider the case of an insertion of a fermionic string coordinate 
$\psi^{\nu}_{(\pm )} (z',\bar{z}')\,$ 
located at the world-sheet boundary. Working out equation 
(\ref{insertion}), but now subject to constraint (\ref{BC}) we find  
\begin{eqnarray}
\label{eqmov2}
< \partial_{z} \psi^{\mu}_{(+)} (z,\bar{z}) \psi^{\nu}_{(+)}
(z',\bar{z}') > &=& \,\pi \,i \, \alpha'\, < \eta^{\mu \nu} 
\delta^2(z-z',\bar{z}-\bar{z}') > \nonumber \\
< \partial_{\bar{z}} \psi^{\mu}_{(-)} (z,\bar{z})
\psi^{\nu}_{(-)}(z',\bar{z}') > &=& \,\pi \,i \, \alpha' \, < \eta^{\mu
\nu}  \delta^2(z-z',\bar{z}-\bar{z}') > \nonumber \\
< \partial_{z} \psi^{\mu}_{(+)} (z,\bar{z})
\psi^{\nu}_{(-)}(z',\bar{z}') > &=&  
\,\pi \,i \, \alpha'\, < \Big[ ( \eta+{\cal B})^{^{-1}}\,
( \eta - {\cal B}) \Big]^{\nu \mu} 
\delta^2(z-z',\bar{z}-\bar{z}') > \nonumber \\
< \partial_{\bar{z}} \psi^{\mu}_{(-)} (z,\bar{z})
\psi^{\nu}_{(+)}(z',\bar{z}') > &=&   
\,\pi \,i \, \alpha'\, < \Big[ ( \eta - {\cal B} )^{^{-1}}\,
(\eta + {\cal B}) \Big]^{\nu \mu} 
\delta^2(z-z',\bar{z}-\bar{z}') > \nonumber \\
\,\,.
\end{eqnarray}

So the appropriate normal ordering for fermionic string coordinates at the boundary 
is
\begin{eqnarray}\label{bno}
: \psi^{\mu}_{(+)} (z,\bar{z}) \psi^{\nu}_{(+)}
(z',\bar{z}') : &=&  \psi^{\mu}_{(+)} (z,\bar{z}) \psi^{\nu}_{(+)}
(z',\bar{z}')  \,-\, \frac{ i \alpha'}{2 (\bar{z}-\bar{z}')} \eta^{\mu\nu}
\nonumber\\
: \psi^{\mu}_{(-)} (z,\bar{z}) \psi^{\nu}_{(-)}
(z',\bar{z}') : &=& \psi^{\mu}_{(-)} (z,\bar{z}) \psi^{\nu}_{(-)}
(z',\bar{z}')  \,-\, \frac{ i \alpha'}{2 ( z - z')} \eta^{\mu\nu}
\nonumber\\
:  \psi^{\mu}_{(+)} (z,\bar{z}) \psi^{\nu}_{(-)}
(z',\bar{z}')  : &=&   \psi^{\mu}_{(+)} (z,\bar{z}) \psi^{\nu}_{(-)}
(z',\bar{z}') \,-\,\frac{ i \alpha' 
\, \Big[ ( \eta + {\cal B})^{^{-1}}\,( \eta  - 
{\cal B}) \Big]^{\nu \mu}}{2 (\bar{z}-\bar{z}')} 
\nonumber\\
:  \psi^{\mu}_{(-)} (z,\bar{z}) \psi^{\nu}_{(+)}
(z',\bar{z}')  : &=&  \psi^{\mu}_{(-)} (z,\bar{z}) \psi^{\nu}_{(+)}
(z',\bar{z}') \,-\,\frac{ i \alpha' 
\, \Big[ ( \eta + {\cal B})^{^{-1}}\,(\eta  - {\cal B}) \Big]^{\nu \mu} }{2 ( z - z')}
\nonumber\\
\end{eqnarray}

These are new results. We found normal orderings that incorporate the effect of 
boundary conditions.  
We see that even in the absence of the antisymmetric tensor background ${\cal B}$
the normal ordering is discontinuous at the boundary. The products
$ : \psi^{\mu}_{(+)} \psi^{\nu}_{(+)}:$ and $: \psi^{\mu}_{(-)} \psi^{\nu}_{(-)}:\,$
are reduced by a factor $\,1/2\,$ on the boundary.
Normal ordered products are important to calculate the central charge that gives us 
the critical dimension. Let us see what happens on the boundary.

The Energy Momentum tensor for the fermionic sector for points inside the 
world-sheet reads 
\begin{eqnarray}
\label{EMT}
T^{zz} &=& - \,\frac{1}{2} \,\psi_{\mu (+)} E^{\nu \mu} \partial_{\bar{z}} \psi_{\nu (+)} 
\,\equiv\,\, {\bar T} \nonumber\\
T^{{\bar z}{\bar z}} &=& 
- \,\frac{1}{2} \,\psi_{\mu (-)} E^{\nu \mu} \partial_{\bar{z}} \psi_{\nu (-)} \,
\equiv \,\, T \,\,,
\end{eqnarray}

\noindent while for at the boundary, 
the conditions (\ref{BC}) relating $\psi_{\nu (-)}\,$ to
 $\psi_{\nu (+)}\,$ lead to  
\begin{equation}
\label{BEMT}
{\bar T} \,= \, - \,\frac{1}{2} \,\,\psi_{\mu (+)} E^{\nu \mu} 
( \partial_{\bar{z}}\,+\,\partial_{z} )\,\psi_{\nu (+)} 
\,=\,  T \,\,.
\end{equation}

The central charge can be calculated from the most singular term in the normal ordered
product of energy momentum tensors. This term involves two contractions of 
Fermionic coordinate operator products and is proportional to

\begin{eqnarray}
\label{central}
\int dz^{\prime}...\int dz^{\prime\prime\prime\prime}& &\cdot\,
\Big[ \frac{- i\alpha'}{2 a\,(z^\prime \,-\,z^{\prime\prime})} 
\,\frac{\delta}{\delta \psi_{\mu (-)} (z^\prime)} 
\,\frac{\delta}{\delta \psi^{\mu}_{(-)} (z^{\prime\prime)}} \Big]
\nonumber\\
& & \cdot\,\Big[ \,\frac{-i\alpha'}{2 a\,(z^{\prime\prime\prime} \,-\,
z^{\prime\prime\prime\prime})}
\,\frac{\delta}{\delta \psi_{\nu (-)} (z^{\prime\prime\prime})} 
\,\frac{\delta}{\delta \psi^{\nu}_{(-)} (z^{\prime\prime\prime\prime)}} \Big]\,
\,\Big[ T (z_1) \,T (z_2 ) \,\Big]\,\,. 
\end{eqnarray}

The contributions involving the antisymmetric tensor $ {\cal B}_{\mu\nu}\,$  cancel 
out. For points inside the world-sheet $a=1$  and the energy momentum tensor $T$ 
is given by eq. (\ref{EMT}).  As it is well known, the central charge is 
$D/2$ where $D$ is the space-time dimension.

For points on the boundary $ a= 2$ as a consequence of the normal ordering 
of eq. (\ref{bno}) but the energy momentum tensor takes the form (\ref{BEMT}).
After taking the functional field derivatives in (\ref{central}) and considering the 
points $z_1$ and $z_2$ to be fixed on the boundary, where $dz = d\bar z \,$, 
the derivatives will act as $\partial_{\bar{z}}\,=\,\partial_{z} \,$.
So, each energy momentum tensor will contribute with a factor of 2 with respect to
the case of points inside the world-sheet.  
As a result, the contribution to the central charge will have, as expected, 
the same value $D/2$ for world-sheet boundary points. 

It is interesting to compare the present fermionic results with those for the bosonic 
string coordinates. In the bosonic case, the normal ordering at the boundary 
reads\cite{Braga:2004wr}
\begin{eqnarray}
\label{no}
\mbox{{\bf :}} \, {\hat X}^{\mu}(w, \bar w )\, {\hat X}^{\nu}(w^\prime, {\bar w}^\prime)
\,\mbox{{\bf :}} &=&
 {\hat X}^{\mu}(w, \bar w ) \, {\hat X}^{\nu}(w^\prime, {\bar w}^\prime) 
+ \frac{\alpha'}{2}
\eta^{\mu \nu} ln \vert w-w' \vert^2  \nonumber\\
&+& \frac{\alpha'}{2}
\Big( [ \eta + {\cal B} ]^{-1}\,[ \eta - {\cal B} ] \Big)^{\mu \nu}\,\, ln (w- \bar{w}')
\nonumber\\ 
&+&  \frac{\alpha'}{2} \Big( [ \eta  + {\cal B} ] \, [ \eta - {\cal B} ]^{-1} 
\,\Big)^{\mu \nu} \,\, ln (\bar{w}- w' )\,+\,
\alpha' D^{\mu\nu}\,\,.\nonumber\\
\end{eqnarray}

\noindent The effect of the boundary conditions is the presence of the terms involving
the $\, {\cal B}\,$ field in this expression. 
They are additive factors that do not contribute to the central charge. 
In contrast to the fermionic case studied here, the bosonic normal ordering 
does not have a discontinuity on the boundary.
So, this is a more trivial situation than the fermionic case where there 
are multiplicative factors coming from the normal ordering 
that cancel with others coming from the energy momentum tensors.

\bigskip
\noindent Acknowledgments: The authors are partially supported by  CNPq, 
FAPERJ ,CLAF and FAPESP.

\end{document}